# Structure and optical properties of Cd substituted ZnO ($Zn_{1-x}Cd_xO$) nanostructures synthesized by high pressure solution route

**Manoranjan Ghosh [1], and A.K.Raychaudhuri[2]**

DST Unit for Nanoscience, S.N. Bose National Centre for Basic Sciences

Block- JD, Sector-III, Salt Lake, Kolkata - 700 098, INDIA.

[1] email: mghosh@bose.res.in

[2] email: arup@bose.res.in, also at the Department of Physics, Indian Institute of Science, Bangalore –560012, INDIA

**Abstract.** We report synthesis of Cd substituted ZnO nanostructures ($Zn_{1-x}Cd_xO$ with x upto $\approx$ .09) by high pressure solution growth method. The synthesized nanostructures comprise of nanocrystals that are both particles (~ 10-15 nm) and rods which grow along (002) direction as established by Transmission electron microscope (TEM) and X-ray diffraction (XRD) analysis. Rietveld analysis of the XRD data shows monotonous increase of the unit cell volume with the increase of Cd concentration. The optical absorption as well as the photoluminescence (PL) shows red shift on Cd substitution. The line width of the PL spectrum is related to the strain inhomogenity and it peaks in the region where the CdO phase separates from the $Zn_{1-x}Cd_xO$ nanostructures. The time resolved photoemission showed a long lived (~10ns) component. We propose that the PL behavior of the $Zn_{1-x}Cd_xO$ is dominated by strain in the sample with the redshift of the PL linked to the expansion of the unit cell volume on Cd substitution.

## 1. Introduction

In recent years there have been renewed interests in studying optical properties of ZnO nanocrystals. It is a wide band gap (3.37 eV) semiconductor with large excitonic binding energy (59 meV) [1]. It can be used as UV lasing material by stimulated emission of exciton at temperature well above the room temperature [2]. For controlled applications, band gap engineering is technologically very important. It is possible to tune the band gap of ZnO by alloying with suitable candidates. Isovalent substitution of Mg leads to enhancement of its band gap [3-5], whereas Cd incorporation reduces the same [6]. In nanocrystals the band gap tuning can also be done by utilizing quantum confinements which leads to blue shift. The band gap engineering leads to a tuning of the PL emission energy and also a change of the life time of the emission. In the context of doped ZnO nanocrystals these issues though studied in the past have not been well understood.

In this paper we describe the synthesis of $Zn_{1-x}Cd_xO$ nanostructures by an energy efficient high pressure solution route which can produce large amount of the $Zn_{1-x}Cd_xO$ nanocrystals of good structural and optical qualities. The nanocrystals tend to have the shape of nanorod .We show that such nanocrystals have near band edge PL that is red shifted significantly and have relatively large PL decay time (~10ns) as revealed through time resolved PL measurements .The nanocrystals do not show the defect related green emission often seen in ZnO nanostructers. We could also link the tuning of the band gap and the PL to expansion of the unit cell as a result of the Cd substitution. The full-width at half maximum (FWHM) of the near band edge emission band has been found to be linked to the strain inhomogenity arising from the process of substitution.

The major challenge faced in the solution route synthesis process is that the CdO is formed at relatively high temperature ($230^0$ C) which is much higher than the boiling point of most of the solvents generally used for solution synthesis. Also because of different crystal structures of ZnO (wurtzite) and CdO (rock salt), they tend to get phase separated very easily. To overcome these problems, the $Zn_{1-x}Cd_xO$ has been synthesized in an autoclaves at high pressure (≈55 atm.) which can increase the boiling point of solvents like ethanol which allow relatively higher temperature synthesis. At this high temperature and pressure, $Zn_{1-x}Cd_xO$ nanostructures form along with CdO. The nanostructures formed consist of both particles and as well as rods of good crystallinity and optical quality

It is noted that Cd substituted ZnO has been prepared so far by vapour phase methods. Epitaxial films of alloyed $Zn_{1-x}Cd_xO$ grown on a substrate have been mostly fabricated by pulsed laser deposition (PLD) [6], molecular beam epitaxy (MBE) [7] and remote plasma enhanced metallorganic chemical vapor deposition (RPE-MOCVD) [8]. Recently $Zn_{1-x}Cd_xO$ nanorods have been fabricated by the thermal evaporation of precursors containing Zn and Cd [9]. Production of nanocrystals of Cd substituted ZnO by vapour phase methods has its limitations because it is expensive and materials obtained are in small quantities. Thus the synthesis of Cd substituted ZnO nanostructures is a very desirable step. There are a large number of reports of synthesis of undoped ZnO by solution or hydrothermal method [10-13]. *Undoped* ZnO nanorods have been synthesized by hydrothermal method very recently [14]. However there has been no published report so far on solution route synthesis of $Zn_{1-x}Cd_xO$ nanostructures. Demonstration that one can produce band gap tuned good quality nanocrystals of Cd substituted ZnO by a low temperature solution route opens up the possibility of large scale synthesis of this material.

## 2. Experimental details

*2.1 Synthesis of nanostructures*

ZnO and $Zn_{1-x}Cd_xO$ nanostructures are synthesized by a simple solution route using a mixture of cadmium acetate and zinc acetate in ethanol. For the synthesis of undoped ZnO nanostructures, 15 mM solution of zinc acetate dihydrate and 45 mM solution of sodium hydroxide are prepared in ethanol. The solution is sonicated until it becomes clear. The clear solution thus prepared is taken in a Teflon lined autoclave and kept inside an oven preset at 230 $^0$C. The ZnO nanostructures of average sizes varying from 15-20 nm are formed after two hour reaction within the autoclaves. The nanostructure thus fabricated is sonicated for half an hour to get the proper size dispersion. To obtain the $Zn_{1-x}Cd_xO$ nanostructures, an ethanol based solution of Zinc acetate dihydrate, cadmium acetate dihydrate and sodium hydroxide was prepared and similar steps as described above have been followed. The dispersed solution contains two phases: the hexagonal $Zn_{1-x}Cd_xO$ of average size 15-20 nm and CdO of average size 70-80 nm having rock salt structure. Precipitation of only CdO particles is possible by centrifuging the solution at 2000 r.p.m. for one minute. The filtrate containing $Zn_{1-x}Cd_xO$ is again centrifuged at 5000 r.p.m. for four minutes to get the precipitation. The $Zn_{1-x}Cd_xO$ precipitate is again dispersed in ethanol and centrifuged for several times to wash the material.

*2.2 Structural and Optical measurements*

The X ray diffraction (XRD) as well as the Inductively Coupled Plasma Atomic Absorption Spectroscopy (ICP-AES) were carried out on precipitates dried at 120 $^0$C. The precipitate was re-dispersed in ethanol and taken in a cuvette for optical measurement. A very crucial parameter in the synthesis of Cd substituted ZnO is the determination of the exact Cd content. The Cd content x was determined by ICP-AES as well as by Energy dispersed X Ray (EDX) analysis. However, we found that the EDX analysis shows approximately 50 % lower Cd content as compared to the ICP-AES method. We have considered the ICP-AES test results as a more accurate method due do its quantitative nature of calibration which have been done with the available standards. Nanostructures and their lattice fringes were imaged by a JEOL [15] High resolution Transmission Electron Microscope (HRTEM) working at 200 KeV. The crystal structures were studied by Rietveld analysis of the XRD data obtained using a Philips Xpert-Pro [16] X-ray diffractometer. Combination of XRD and the TEM studies helped us to analyze the structure when Cd is progressively alloyed in ZnO.

Room temperature optical properties were studied using a UV-visible spectrophotometer (Shimadzu UV-2450 [17] and fluorimeter (JOBIN YBON – Fluoromax-3 [18] which uses a Xenon arc lamp as the illuminating source. The excitation wavelength chosen for the PL measurement is 325 nm. We carry out Time resolved photoluminescence (TRPL) measurements on the Cd doped nanostructers. For time resolved measurement samples were excited at 299 nm, by a pulsed light emitting diode of 40 MHz repetition rate. The PL decay time were measured by using time correlated single photon counting technique [19]

## 3. Results

*3.1 Standardization of the method of synthesis and composition analysis*

The synthesis of $Zn_{1-x}Cd_xO$ nanostructures by solution method is challenging because of two reasons: 1) CdO is formed only at temperature above $220^0$- $230^0$ C. At temperature below 100 $^0$C the synthesis by the acetate route leads to only $CdCO_3$ formation. Up to 170-180 $^0$C, $Cd(OH)_2$ is formed and no incorporation of Cd into ZnO takes place. Incorporation of Cd in ZnO occurs only when the growth temperature is above $200^0$C. This necessitates a high temperature solution route. 2) Because of different structures of ZnO (hexagonal) and CdO (rocksalt), they tend to get phase separated easily. Most of the Cd added to the solution leads to formation of CdO and only a small part gets incorporated into the ZnO lattice.

To obtain higher temperatures, the reaction is made to happen within an autoclaves working at 230-240$^0$C. At this temperature range CdO is formed and incorporation of Cd into ZnO is also initiated. The formation of CdO can be detected by visual observation. As soon as CdO formation starts the bluish color of the ZnO dispersion turns reddish. As the Cd concentration is increased in the reactant further the incorporation of Cd into ZnO takes place along with a phase segregation of CdO. The two phases thus formed have different sizes. While the $Zn_{1-x}Cd_xO$ nanocrystals have an average size of ~10-15 nm, the CdO particles have a much larger size ~80 nm. This difference in size allows us to separate the two phases by centrifugation.

The subsequent compositional analysis as well as structural and optical studies were conducted after the two phases were physically separated.

Analysis of the $Zn_{1-x}Cd_xO$ nanocrystals by ICP-AES shows that the maximum observed Cd incorporation into the ZnO nanostructures correspond to $x \approx 0.091$ which is comparable to some of the vapor phase synthesized samples [20]. However there are reports of some of vapor phase grown samples where a very high value of x have been observed [21]. It appears that the measure of exact Cd incorporated in a sample is dependent on the analytical technique used. In figure 1 we show the exact value of 'x' as determined through ICP-AES analysis vs. the molar fraction of $Cd(CH_3COO)_2$, $2H_2O$ added in the $Zn(CH_3COO)_2$, $2H_2O$ solution. As mentioned before, we use the ICP-AES data because it is a bulk analysis method that is based on quantitative standards. We feel that the difference in the Cd concentration analyzed by the two techniques (EDX and ICP-AES) arises because while ICP-AES technique is averaged over the bulk, the EDX is a surface sensitive technique being limited by the electron penetration depth and what it reports is essentially the concentration near the surface. Due to the segregation of the CdO out from the nanocrystals, there may be a concentration gradient which increases from the interior of the nanocrystals to the surface and the difference in the Cd concentration seen by the two techniques may reflect this. The quantitative compositional analysis (irrespective of the technique used) establishes that the Cd incorporation is indeed very low due to the high lattice mismatch and difference in crystal structure of ZnO and CdO. We will see below that the incorporation Cd into ZnO gets limited because phase segregation occur very early ($x \approx .03$) in the growth process where the $Zn_{1-x}Cd_xO$ nanocrystals physically separate out from the CdO crystals those have relatively larger size.

*3.2 Characterization by Transmission electron microscope*

The size and microstructure of the undoped and doped samples have been investigated in detail by Transmission electron microscope (TEM). For low x ( $x < 0.03$) the nanostructures those are formed are predominantly spherical shaped nanocrystals. As an example we show in figure.2 (a) the particle size distribution of the undoped ZnO sample. The nanostructures are predominantly spherical in shape and the average sizes ~15-20 nm. This is similar to that found in the synthesis route under normal pressure and at temperature $70^0$C [5]. For particles with low values of x, the size distributions are very similar to the undoped nanocrystals. The lattice image (figure.2 (b)) confirms the crystalline nature of the sample and the distances between two consecutive lattice fringes (2.41 Å) is similar to the interplanar spacing of the (101) set of planes as indicated in the figure.

      The spherical shape of the nanostructures no longer persists as soon as significant Cd incorporation starts. This is shown in figure 3. For $x \geq 0.03$, it can be seen that along with spherical nanocrystals one also sees formation of hexagonal nanocrystals as well as the formation of nanorods as marked in the figure. The nanocrystals have typical size from 15-20 nm (x < .03) to 10 nm for x >.05. The rods have typical diameter of ~ 10-12 nm and length 20-25 nm. The histogram of the aspect ratio is shown in the inset of figure 3. The trend of rod formation persists till higher concentration of Cd and eventually the tendency for rod formation predominates. In figure.4 (c) we show the HR-TEM image of a single rod (formed for x=0.065) which is aligned in the (002) direction. The lattice spacing (2.68 Å) matches with the interplanar spacing of the (002) set of planes. As shown in figure 4 (d), the FFT (Fast Fourier Transform) of the lattice image confirms that the rods are single crystalline and have only (002) orientation as the growth direction. This preferred orientation also has been confirmed by the XRD data analysis, presented later on. For comparison in figure 4 (a) and (b) we show the High Resolution TEM and its FFT of a spherical nanocrystal respectively which shows hexagonal symmetry of the lattice. At higher concentration rods are synthesized in higher proportion but they retain their size and aspect ratio as well as the orientation (002). The single phase nature of the $Zn_{1-x}Cd_xO$ is confirmed by TEM as well as by the XRD data for $x \leq .091$. There is no further incorporation of Cd in ZnO beyond this value of x. Addition of more Cd salt in the solution leads to formation of more CdO.

*3.3 Structural analysis by X-ray diffraction data*

The structural analysis of the powdered X ray diffraction data of the $Zn_{1-x}Cd_xO$ nanostructures has been carried out extensively by using Rietveld method [22]. The XRD data analysis also allows us to establish the range of Cd concentration where the phase segregation occurs. We could go to a maximum Cd incorporation of x=0.091 as stated before. The XRD patterns of the $Zn_{1-x}Cd_xO$ alloy nanostructures are shown in figure.5. The nanocrystals have wurtzite structure (peaks are indexed) as that of the parent ZnO (x = 0). However the peak positions shift to lower $2\theta$ values due to the Cd incorporation signifying change (increase) in the lattice constant. (Note: We find from the XRD data also that a CdO phase appears at a rather a low level of x=.03. However, as mentioned before most of the Cd forms the CdO phase with nanoparticles with average size of ~ 70-80 nm. We have separated out the CdO particles by centrifuging and the XRD data shown are only from the $Zn_{1-x}Cd_xO$ nanostructures with average size ~ 10-20 nm which are single phase.) The XRD data along with the TEM data presented before establish that the high pressure solution route can be used to synthesize nanocrystals of single phase $Zn_{1-x}Cd_xO$ alloy of good crystallinity.

The texture growth of rod in the nanostructure synthesis as the Cd content is increased, as seen in the TEM images, is also clearly seen in the XRD data. The relative intensity of the Bragg peak due to (002) planes of the wurtzite structure increases with the increase in Cd content. This fact can be clearly seen from Figure. 6, where the intensity ratios of the Bragg peaks (100) and (002) with maximum intense peak (101) of the wurtzite phase has been plotted with the increase in Cd content. The important observation is that the intensity ratio (002)/(101) increases sharply up to x= 0.03, beyond which it saturates. To establish that the growth occurs preferentially in the (002) direction, we have plotted for comparison the intensity ratios of (100)/(101) peaks which has a more or less constant value It confirms the increase in the preferred growth of rods aligned in (002) direction as we increase the Cd concentration.. The (002) direction is an easy axis of growth for ZnO and most of the nanorod formation in ZnO occurs along this orientation.

There is yet another important observation that accompanies the process of texturing. The XRD peaks of the wurtzite phase show enhanced full-width at half maximum (FWHM) values for higher Cd content. We have carried out a Williamson- Hall analysis [23] of the XRD data. The analysis show that the increase in the XRD line width occurs due to an enhancement of the inhomogeneous strain (which we refer to as the microstrain) as well as due to a decrease in the average size of the $Zn_{1-x}Cd_xO$ nanostructures with the increase in Cd content. The enhancement of the microstrain, as seen through the XRD analysis also widens the FWHM of the PL peak as we see later on. Observed increase in strain values can be due to the effect of fluctuation in the Cd concentration as previously suggested by Makino et al. [6].

Variation of the lattice constants and crystallite sizes were obtained from Rietveld analysis of the observed XRD data. As the Cd content is increased, the peak positions of the XRD data shift to lower angles due to increase in lattice volume which is caused by the incorporation of $Cd^{++}$ ion having higher ionic radii (0.97 Å) in place of the $Zn^{++}$ ion (0.74 Å) in the ZnO lattice [24]. The calculated profile matches well with the observed data as shown in Figure 5. In this least square refinement the peak shape is assumed to be pseudo-voigt. The residues of the fitting for all the data are indicated in figure 5. We have noted before (as measured through the HR-TEM observation) that enhanced incorporation of Cd in ZnO also changes the growth morphology. We could find signature of this in the analysis of the XRD data. We find that the average size of the $Zn_{1-x}Cd_xO$ nanostructures observed to be decreased from 15-20 nm (x = 0) to 10 nm for x >.05. Beyond that the size remains constant.

The lattice parameters ('c' and 'a' axes) extracted by Rietveld analysis show that they increase monotonously as the Cd content is increased. The change is small but it is distinct. In Figure 7 we show the variation of the unit cell volume ($V=0.866a^2c$), the microstrain as well as the *c/a* ratio as function of the Cd content. The cell volume increases by nearly 0.8% on Cd substitution. It is interesting to note that the small yet distinct compaction of the lattice does not preserve the c/a ratio. The most of the change occurs in the a-axis which increases by 0.3 %, while the c-axis lattice parameter expands only by 0.19%. This reduces the *c/a* ratio. Here we note that the expansion of the unit cell volume on Cd substitution can be thought of as a "negative pressure" arising from the larger Cd ions. The negative pressure is not hydrostatic and leads to preferential compaction of the a-axis leading to a smaller c/a ratio. Increase of the unit cell volume on Cd substitution has also been found in films grown by vapor phase [6].

From figure 7 it appears that there are two stages in the Cd substitution. The change over occurs at around x ≈ 0.03. Below x=0.03 the nanocrystals are predominantly spherical. In this region the microstrain increases till the CdO separates out which occurs beyond x=0.03. Above x=0.03 when the CdO separates out, the growth leads to nanorod formation that have preferential orientation along (002). The intensity ratio (002)/(101) shown in figure 6 reflects this change of stage at x. ≈ 0.03. The intensity ratio saturates beyond x=0.03 indicating preferential growth of (002) oriented nanorods. Figure 7 also indicates that beyond x=0.03, the phase separation releases some of the microstrain in the lattice.

The detailed structural investigations presented before have established that single crystalline nanocrystals as well as rods of well characterized structural quality can be synthesized by the high pressure solution route. The structural studies have also established some of the subtle features in growth of Cd substituted nanocrystals, in particular the stages of growth as well as the rod of the textured structure are interesting observation that have not been reported before. There is a continuous expansion of the unit cell volume on Cd substitution with a small reduction in the c/a ratio for the lattice constants. Below we present the results of the optical studies done on these nanocrystals.

*3.4 Optical properties*

Investigation of the optical properties of the chemically synthesized alloy $Zn_{1-x}Cd_xO$ nanostructures is a principal motive of the present investigation. The structural studies presented above show the quality of the material grown and also reflect the small yet distinct changes in the crystal lattice that are likely to have an impact on the optical properties. We have investigated the change in the absorption, the Photoluminescence (PL) and also the time resolved PL (TRPL) in these materials at room temperature. These measurements allow us to assess the utility of these chemically synthesized materials for application as a band gap engineered optical material.

*(a) UV-visible absorption* - One of the important effects of Cd incorporation into ZnO is the reduction of its direct band gap value as has been seen in vapor phase synthesized $Zn_{1-x}Cd_xO$ nanostructures. The band gaps of the alloy nanostructures were determined by monitoring the onset of the fundamental absorption edge. As shown in figure 8, a sharp peak appears in the absorption spectra of undoped ZnO nanostructures even at room temperature due to the large excitonic binding energy (59 meV) of ZnO [1]. The absorption curves shift to the lower energy for the alloy nanostructures. The absorption edge can be clearly identified from the optical absorption upto $x < 0.03$. Beyond that the absorption edge gets blurred probably due to appearance of large number of disordered induced states within the band gap and partially due to the CdO present along with the $Zn_{1-x}Cd_xO$ nanostructures. The shift in the band edge (for $x < 0.3$) as a function of x are shown in figure 10. The direct band gap value of the undoped ZnO nanocrystals (~10-15 nm) is found to be 3.37 eV which is the fundamental absorption edge of the bulk ZnO. In the size regime we are working, no quantum confinement effect has been observed [25]. One can see a monotonous decrease of the gap with increase of the Cd content. ( Note: Although we have observed decrease in the average size of the nanostructures on Cd incorporation, the size is still larger than the size range where quantum confinement effects (diameter < 5nm) can alter the band gap. Even if there is any effect of quantum confinement, this will reduce the extent of red shift.). Thus the limited absorption data can be analyzed to establish that the absorption edge of the chemically synthesized $Zn_{1-x}Cd_xO$ nanostructures also can be red shifted clearly as in the vapor phase grown materials and the shifts are comparable [20].

*(b) Photoluminescence* –A more clearer signature of the red shift on Cd substitution can be seen from the photoluminescence data. The photoluminescence (PL) spectra of pure ZnO and $Zn_{1-}$ $_x$Cd$_x$O nanostructures have been measured with an excitation of 325 nm. The emission spectra following the 325 nm is shown in figure 9. The undoped ZnO nanostructures show sharp emission in the UV region (P$_1$) at 371 nm (3.345 eV). The emission at 371nm has been established due to exciton recombination. A broad and weak emission (P$_2$) is also observed at 389 nm (3.189eV), the exact nature of which has not been identified but it arises also from excitons those are bound to defects [26]. We do not see any green emission at ~450-500 nm which has been seen in a number of nanoparticles of ZnO [5]. The green emission is typically taken as a signature of defects (like oxygen vacancy) in the ZnO [27]. (Note: In most nanocrystals and quantum dots of ZnO the emission at room temperature occurs at around 3.25-3.35eV. While in bulk single crystals at room temperature the UV emission at ~3.3 eV is attributed to free exciton, the UV emission in nanocrystals and Quantum dots are due to acceptor or donor bound excitons that can arise from lattice defects [28, 29].)

The Zn$_{1-x}$Cd$_x$O nanostructures show P$_1$ as in the case of undoped ZnO and there is no significant shift in the position of the component P$_1$ on Cd substitution. However, the relative contribution of this line to the total emission reduces significantly. The main change occurs in the broad peak P$_2$ which show gradual red shift as well as enhanced intensity due to the incorporation of Cd. The gradual red shift of emission energy of P$_2$ with Cd content is shown in figure 10. For the highest Cd content (x = 0.091) P$_2$ occurs at 417 nm (2.976 eV) which is 213 meV red shifted from the undoped ZnO. In figure 10 we also show the intensity ratio of the two peaks (P$_2$/P$_1$) as a function of x. The figure shows the shift in the intensity as the alloying takes place. However, after the first stage of growth (x ≈0.03) the ratio saturates.

Substantial change in the PL characteristics can be seen in the line width of the P$_2$ band. The full width at half maximum (FWHM) of the P$_2$ line changes as the Cd concentration increases (see figure 11). It can be seen that the dependence is not monotonous. It first increases with x reaching a maximum at x ≈ .04-.045 and then decreases again. For the line P$_1$, the FWHM is more or less constant but it shows a shallow peak near the composition range where FWHM of the line $P_2$ also shows a peak. We note here that the line width appears to have a relation to the micostrain whose variation with x is shown in figure 7. The microstrain reaches a maximum for x ≈ .03-.04. We discuss this issue more later on.

*(c) Time Resolved Photoluminescence* – We have taken time resolved photoluminescence (TRPL) data on the nanocrystals of $Zn_{1-x}Cd_xO$. The representative data taken on three compositions are shown in figure 12. The TRPL data have been taken on the line $P_2$ after excitation at 299 nm. The lifetime decay curves have been resolved into three components by deconvolution fitting of the data with the relation:

$$I(t) = \Sigma_{i=1-3} A_i \, exp \, (-t/\tau_i)$$

The parameters $A_i$ *(i=1,2,3)* which give the weight of the components and the decay time constants $\tau_i(i=1,2.3)$ of the transients are shown in table 1. The first observation that we make is that there is actually an increase in all the decay time constants on Cd substitution. The natural decay time of free excitons in ZnO is rather small and in the range of 1 ns. The decay times observed here are larger suggesting that the emission mainly involves bound exciton. The change in the decay time constant on Cd substitution is different for the three components. It is marginal for the longest time constant $\tau_3$ (~10%), noticeable for the intermediate time constant $\tau_2$ (~ 25%) and substantial for the shortest time constant $\tau_1$ which changes from around 260ps to a more than 500ps. We find that the maximum decay occurs through process that has the longest decay time constant $\tau_3$ (≈ 9-10 ns) which has a relative weight of nearly 50% and there is a very little change in its weight on alloying

**4. Discussions**

In this section we discuss the physical implications of some of the important observations that came out from this work. The issues that we discuss are about growth, PL properties and what appears to us as the essential role of the lattice strain on Cd substitution that affects the growth, phase separation as well as the PL property.

*4.1 Growth*

We find that the growth of the nanocrystals change characteristics when the Cd is gradually substituted. The two phases in growth with a change at x ≈ .03 was pointed out before (see subsection 3.2). One of the observations about the growth is that when the Cd concentration increases the average size of the nanocrystals grow to about half of the size obtained in the undoped case and there is formation of nanorods with (002) growth direction that predominates near the higher range of Cd concentration. We propose to explain this observation in the following manner. It is known that in solution phase , if the initial nucleation process is fast , the growth generally occurs by such coarsening process as Oswald ripening (OR) , which is diffusion limited or through the aggregation by orientation (AO) that leads to ordered growth of nanocrystals with different morphology [30]. The growth of nanorods of ZnO as has been observed by us is likely due to the AO process, which is similar to the growth of $TiO_2$ nanocrystals from solution [30]. The occurrence of the AO process as an alternate growth process occurs when the diffusion dependent process like OR or by direct precipitation from solution gets limited. The size limited growth of nanocrystals on Cd substitution is a manifestation of the reduction of such diffusion dependent path ways. Two processes can limit the diffusion dominated processes. The presence of strain on Cd substitution can limit growth as has been observed in a number of crystal growth processes [31, 32]. This occurs due to increase of free energy of the strained nanocrystals. Another alternate reason can be presence of $Cd^{++}$ ions. The incorporation of Cd into ZnO during the growth occurs through absorption of $Cd^{++}$ ions on the surfaces of the growing crystals. It is known that absorption of ions can limit the growth rate in the solution growth process [33]. To summarize this part, the presence of strain and /or adsorption of $Cd^{++}$ ions can limit the growth through diffusion limited process leading to growth by agglomeration by orientation that gives rise to nanorod with (002) orientation.

*4.2 Photoluminescence*

The effects of the Cd substitution are clearly apparent on the PL properties. The main effect is indeed on the emission line $P_2$. The line red shifts and broadens (as revealed through the enhancement of the FWHM). This red shift is substantial and is similar to that found in the vapor phase grown samples. The position of the line $P_2$ appears to be tagged to the red-shift of the band edge with the Stokes' shift nearly remaining constant on Cd substitution. (Note: The data on band edge shift is limited to x < 0.03).

In ZnO crystals it has been shown that a hydrostatic pressure can blue shift the band gap [34]. In films of wide band gap semiconductors like ZnO grown on substrates with lattice constant mismatch, the PL is shifted by the in-plane strain arising due to the mismatch [35]. It has been shown that in films with compressive strain, the PL is blue shifted while tensile strain leads to red shift. We propose that the red shift of the PL in $Zn_{1-x}Cd_xO$ nanostructures, which is a consequence of the red shift of the band edge, arises due to the strain produced by the Cd substitution. The Cd substitution that leads to an expansion of the unit cell can be interpreted as a "negative" pressure producing similar effect as a tensile strain in a film. It is proposed that the "negative" lattice pressure produced by the Cd substitution is the cause of the red shift seen in the PL. We plot in figure 13 the red shift of PL line ($P_2$) as a function of the unit cell size. One can see the smooth variation of the PL line with the unit cell volume.

The line width of the main PL line as measured by the FWHM has a severe sensitivity to the inhomogeneous lattice strain (micro strain) that develops during the process of Cd substitution. As can be seen from figure 7, the microstrain reaches a maximum for x ≈ 0.03-0.04. This is the region where the segregation of the CdO occurs and this phase separation actually reduces the microstarin somewhat. The FWFM of the PL reflects this variation of the microstarin and reaches a shallow maximum for x ≈ .04-.045 (see figure 11). A similar trend, although much weaker is seen the FWHM of the line P1 also suggest that the broadening of the PL line is predominantly due to the strain homogeneity arising due to the Cd substitution.

*4.3 Essential role of strain in the Cd substituted ZnO*

The dependence of the position of the PL line on the cell volume as well as the line width of the PL line on microstrain strongly suggests a link between the emission process and the strain. It has been seen in a number of random substitution alloy semiconductor (e.g., $CdS_{1-x}Se$) [36] that the disorder due to random substitution can lead to smearing of the band edge absorption as well as broadening of the PL line. Its origin is thought to be fluctuations (static) in the band-gap due to random fluctuations in compositions. In addition, the strain field can also lead to exciton localization [37]. The large absorption in the $Zn_{1-x}Cd_xO$ nanostructures in the subband gap region suggests presence of disorder induced states from the random substitution. It is likely that the microstrain seen in the Cd substituted samples arise from the randomness of the substitution process. The Cd substitution sites can provide the excitonic binding sites. This type of strain induced localization can trap the exciton and can lead to large PL decay time as has been seen in this investigation.

**5. Conclusion**

We show that a high pressure solution route can be used to successfully synthesize nanostructures of $Zn_{1-x}Cd_xO$. The maximum Cd incorporated is x ≈ .09. The synthesized nanostructures comprise of nanocrystals that are both hexagonal particles (~ 10-15 nm) and rods which grow along (002) direction. The synthesis of the nanostructure is accompanied by phase segregation of CdO that separates out as large (~70nm-80nm) particles. The Cd incorporation enhances the unit cell volume and the optical absorption edge as well as the photoluminescence emission show red shift. The Cd incorporation is like a "negative pressure" that is similar to a tensile strain. We find a close link of the strain to the PL properties where the microstrain broadens the PL line width. The time resolved photoemission showed a long lived (~10ns) component. We suggest that the long life time arise from exciton localization due to disorder created by the random strain field.
.

**Acknowledgement**

The authors want to thank Department of Science and Technology for financial support as Unit for Nanoscience. Technical support by Mr. Supriya Chakraborti for use of the TEM facility at Indian Association for the Cultivation of Science is acknowledged.

Table 1. Decay constants

| Samples | $\tau_1$ (ps) | $\tau_2$ (ps) | $\tau_3$ (ps) | $A_1$ (%) | $A_2$ (%) | $A_3$ (%) |
|---|---|---|---|---|---|---|
| X = 0 | 266 | 1203 | 8594 | 10 | 38 | 52 |
| X = 0.034 | 532 | 1555 | 9355 | 19 | 29 | 52 |
| X = 0.065 | 544 | 1482 | 9375 | 16 | 30 | 54 |

**Figure Captions**

**Figure 1**. The value Cd content 'x' observed by ICPAES analysis against the mixing of various amount of Cd(CH$_3$COO)$_2$, 4H$_2$O in the Zn(CH$_3$COO)$_2$, 2H$_2$O solution.

**Figure 2.** (a) The average size distribution and (b) the HR-TEM image of undoped ZnO (x = 0) nanostructures.

**Figure 3**. Growth of faceted nanocrystals along with hexagonal nanorods of Zn$_{1-x}$Cd$_x$O (indicated by arrows) for x > 0.03 is shown. The diameter distribution and aspect ratio distribution of the nanorods are depicted in the same figure.

**Figure 4.** High resolution transmission electron microscope of the hexagonal nanoparticle and the nanorod are shown in figure 4. (a) and (c) respectively. The FFT of the lattice fringes of (a) and (c) are shown in (b) and (d) respectively. The growth direction of the single crystalline nanorods ((002)) can be seen.

**Figure 5**. The XRD data for single phase wurtzite Zn$_{1-x}$Cd$_x$O nanostructures. For x = 0.091 the segregated CdO (rock salt structure, peaks are indexed) coexist with Zn$_{1-x}$Cd$_x$O. The profile fitting of the observed data and the residue has been shown against each curve as indicated in the figure. The gradual increase in the relative intensity of the (002)

peak of wurtzite phase can be seen for higher Cd concentration indicating predominance of the (002) aligned nanorods.

**Figure 6**. The intensity ratios of the Bragg peaks (100) and (002) to the maximum intense peak of the wurtzite structure (101) has been plotted with the Cd content x. The enhanced intensity ratio of (002)/(101) peaks indicates increase in the preferred orientation in (002) direction. In comparison the value of (100)/(101) ratio is mainly unchanged as Cd concentration is changed.

**Figure 7.** Unit cell volume, *c/a* ratio and the microstrain as a function of x as obtained form analysis of the XRD data. The arrow shows the composition x=0.03, where a change in stage of growth seems to be occurring.

**Figure 8**. Room temperature absorption spectra show shift to the lower energy values indicating reduction in the direct band gap values as a result of Cd incorporation. Broadening of the excitonic peak occurs as soon as Cd incorporation takes place. For x > 0.03 determination of the band gap is not possible due to broad nature of the absorption.

**Figure 9**. The room temperature emission spectra for x = 0, x = 0.034 and x = 0.091 are shown in Fig (a), (b), and (c) respectively. The $P_2$ emission band of $Zn_{1-x}Cd_xO$ nanostructures show red shift due to the incorporation of Cd as indicated in the figure.

**Figure 10.** The variation of the emission energy of $P_2$ band and the direct band gap values as a function of x. The variation of the relative intensity of the $P_2$ and $P_1$ lines are also shown in the same figure.

**Figure 11.** Variation of the line width of the $P_1$ and the $P_2$ emission lines (Full width at half maximum ) as a function of x.

**Figure 12.** Time resolved decay curves of the PL for three different values of x as indicated in the figure.

**Figure 13.** Variation of the $P_2$ emission line as a function of the unit cell volume as determined from the XRD data**.**

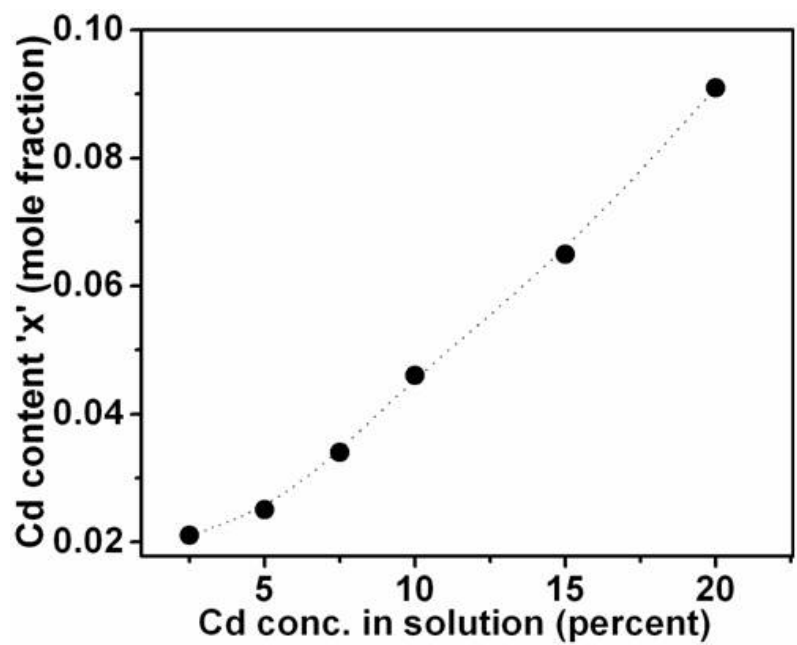

Figure 1

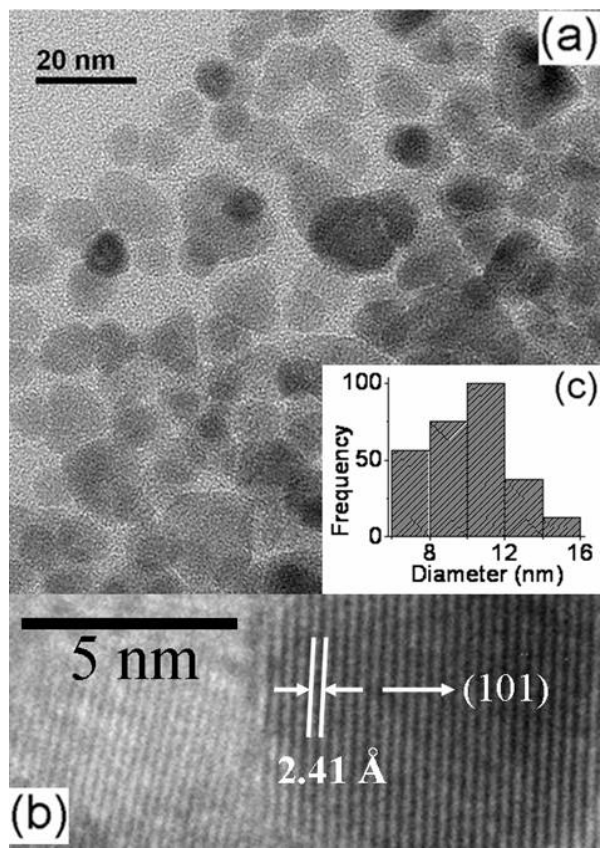

Figure 2

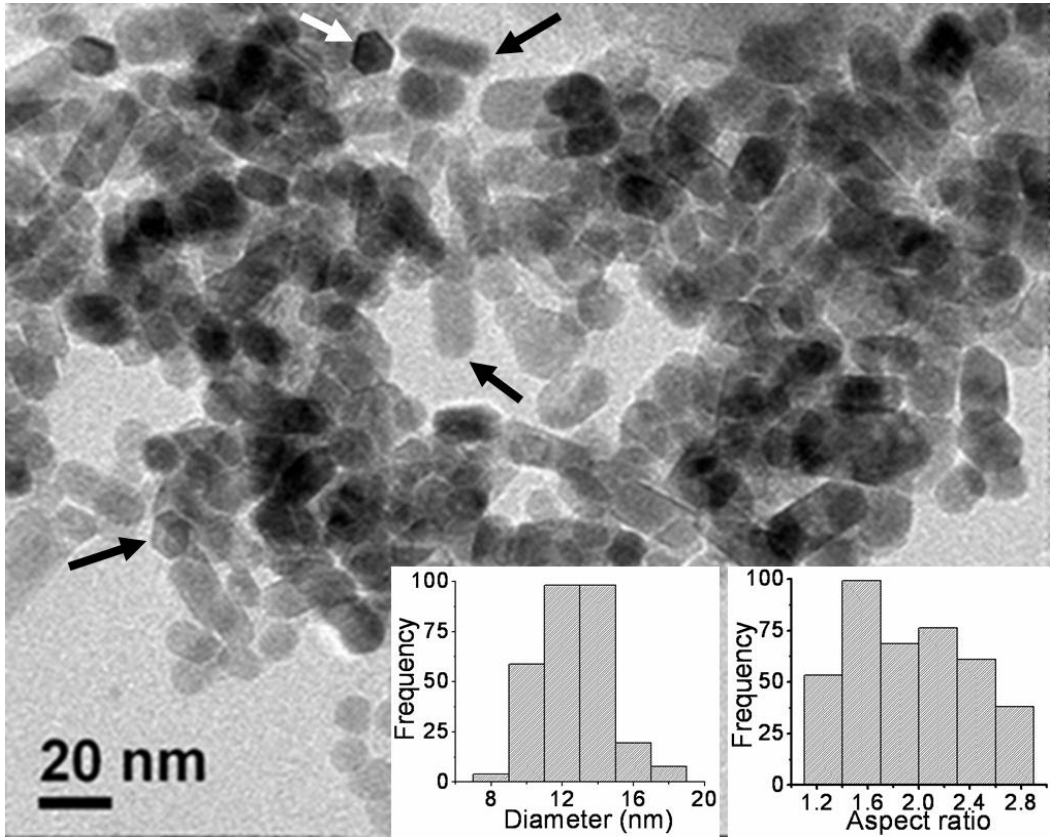

Figure 3

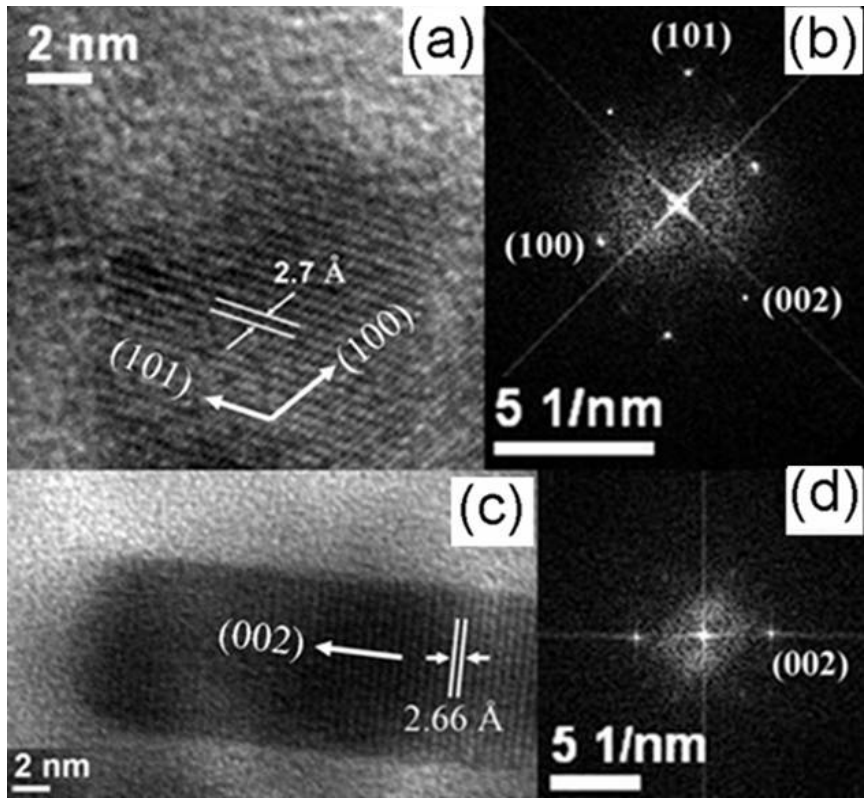

Figure 4

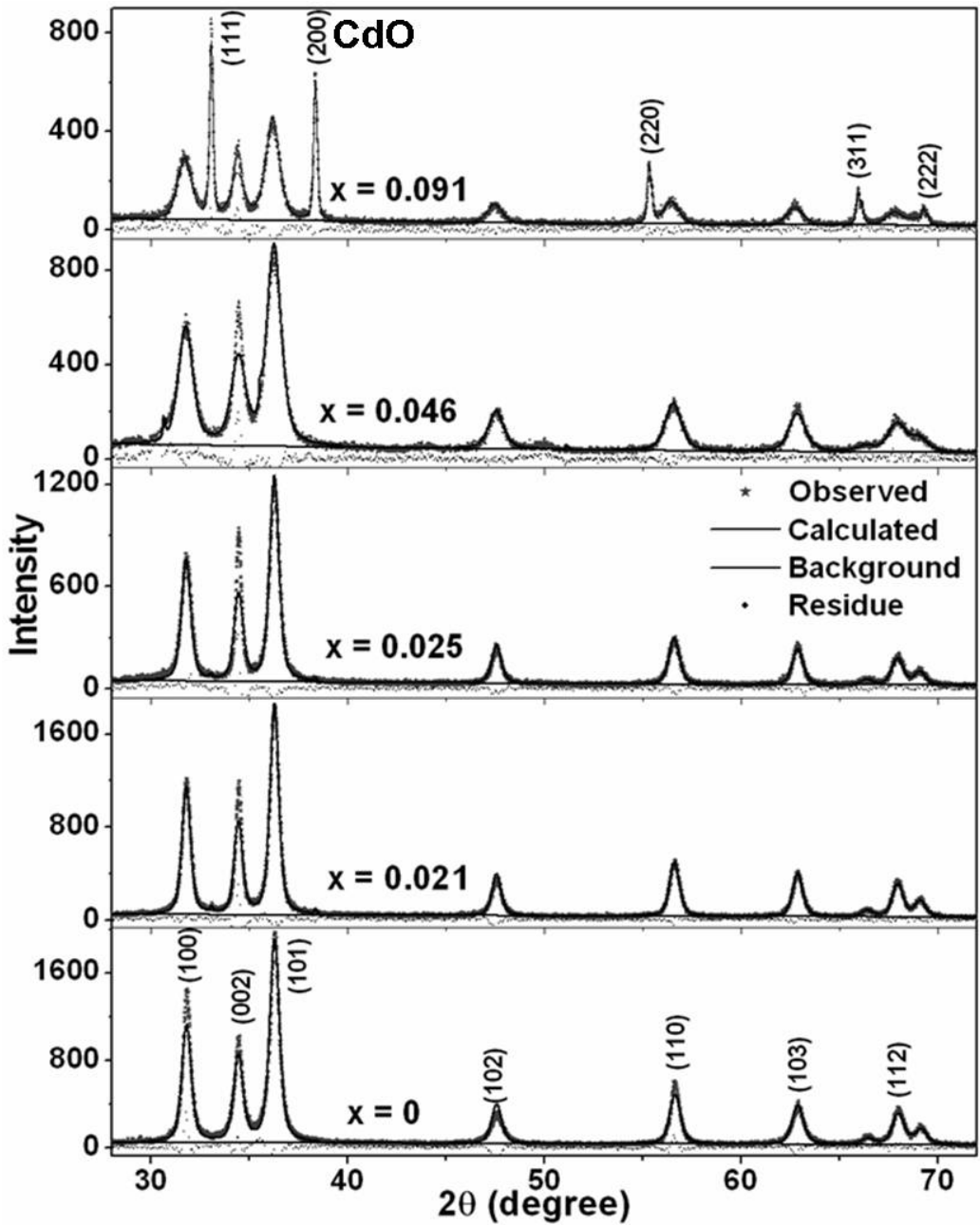

Figure 5

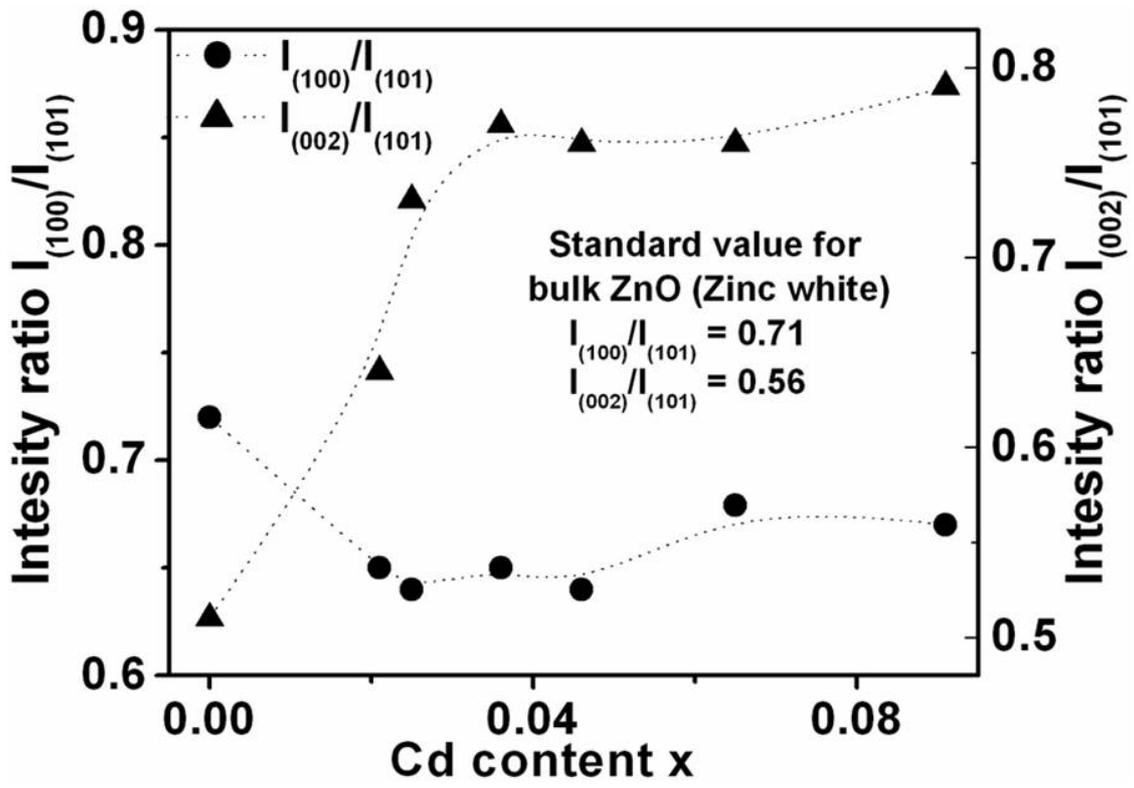

Figure 6

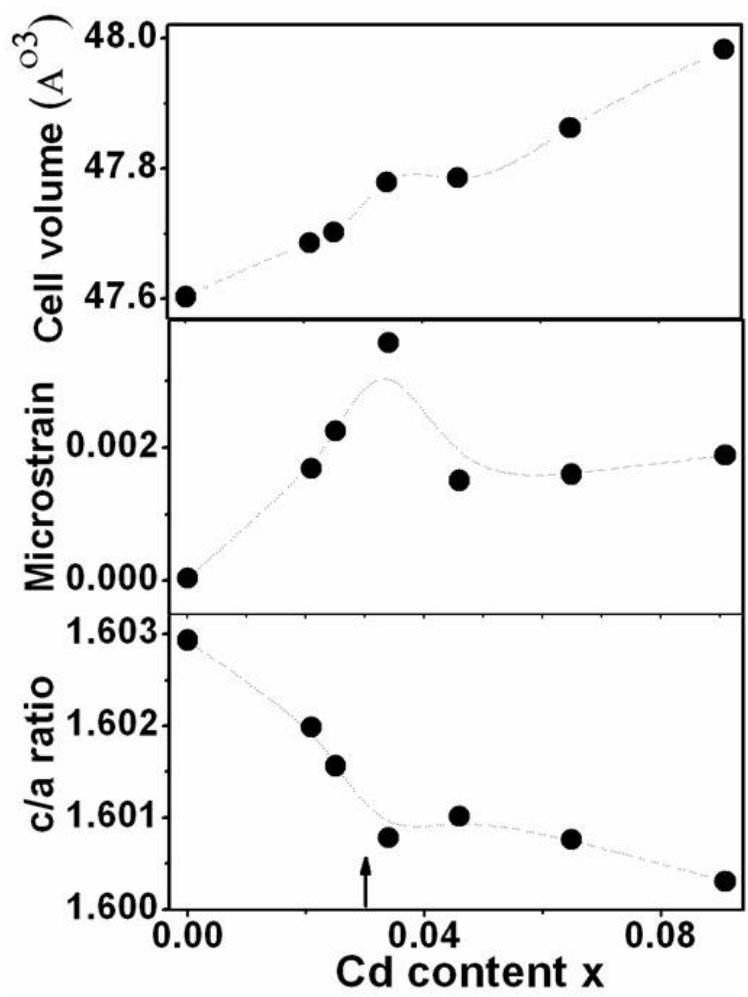

Figure 7

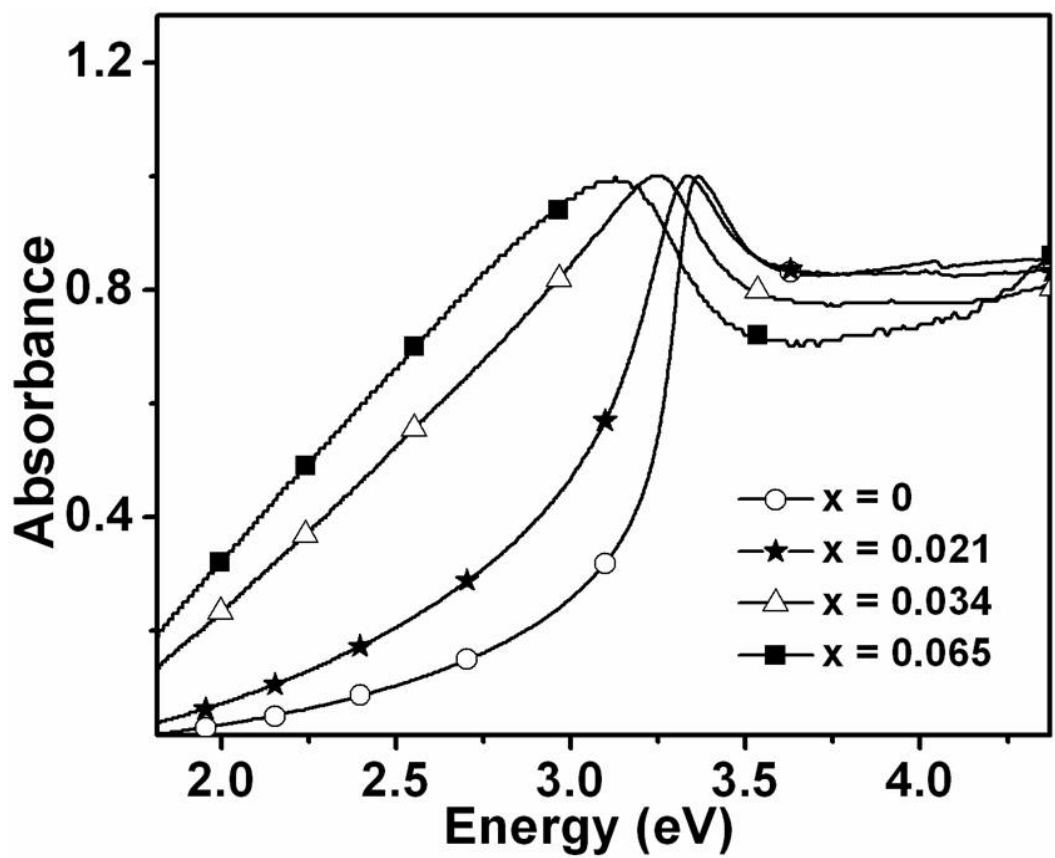

Figure 8

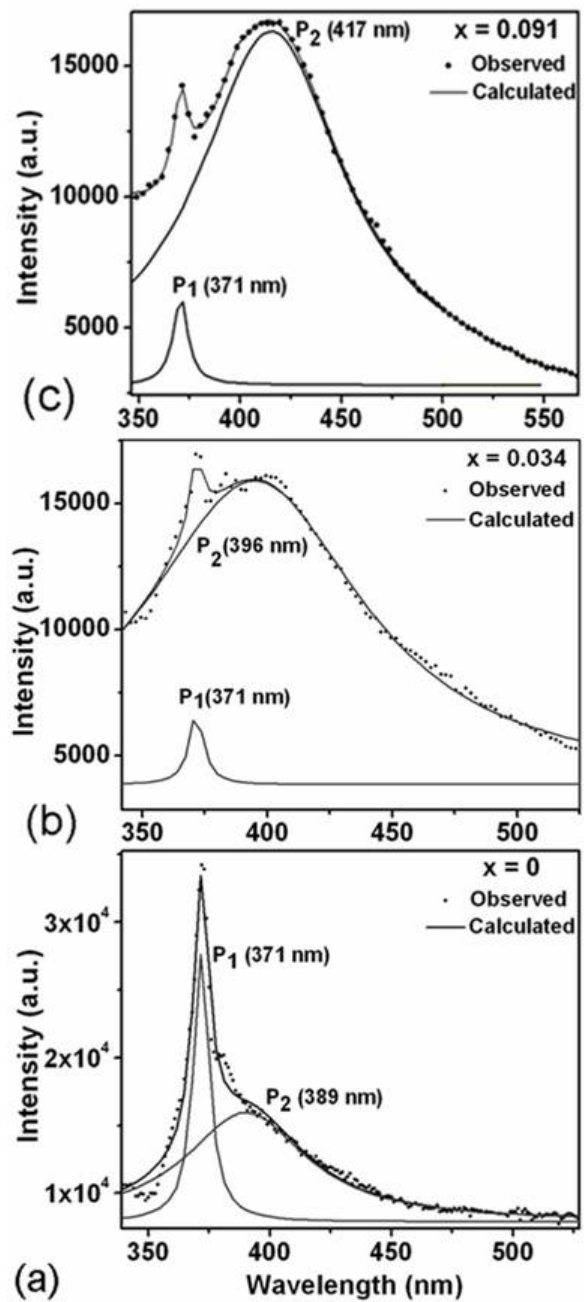

Figure 9

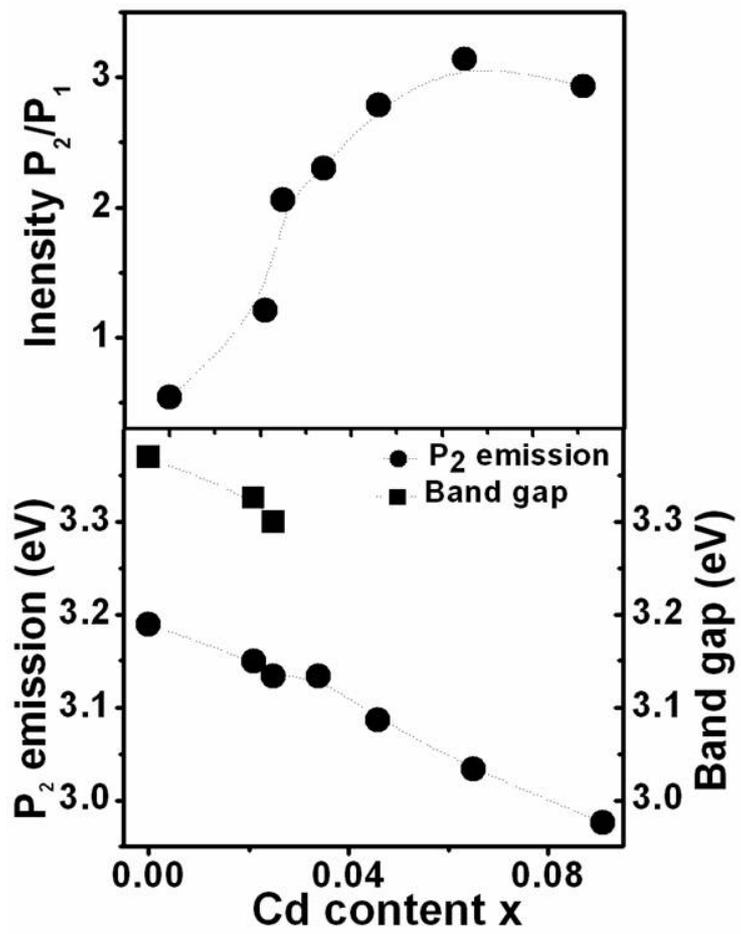

Figure 10

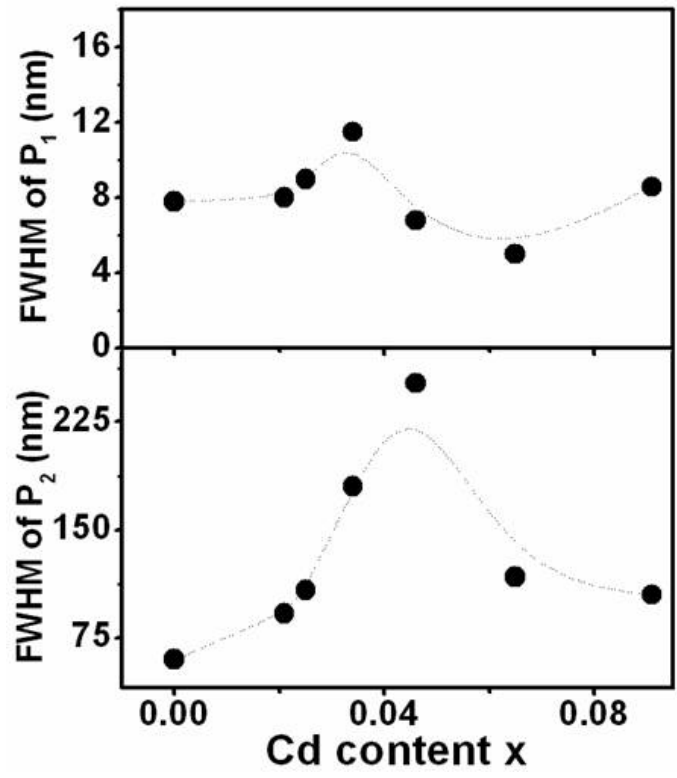

Figure 11

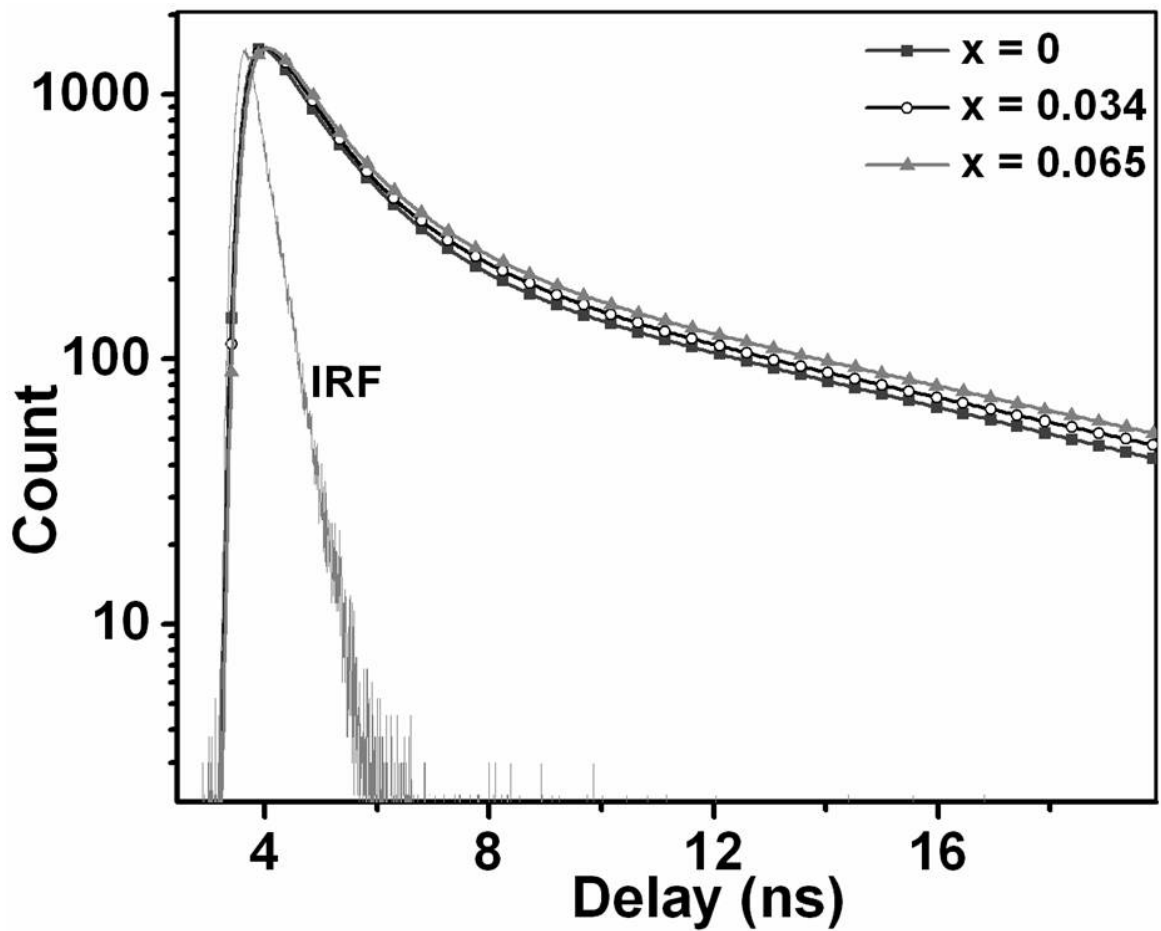

Figure 12

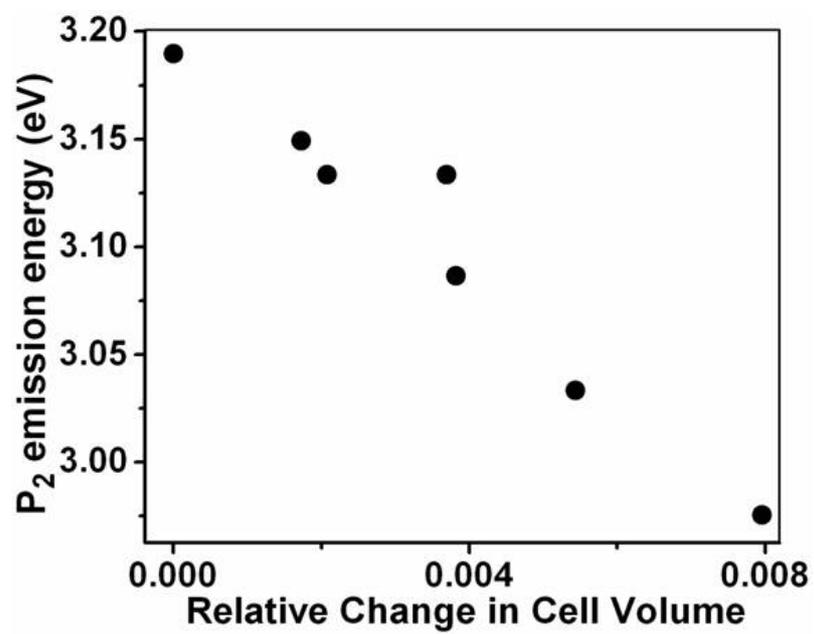

Figure 13